\documentclass[twocolumn,RNAAS]{aastex62}
\graphicspath{{./}{figures/}}

\submitjournal{RNAAS}

\shorttitle{LSST Jupiter Analogs}
\shortauthors{Buzasi}

\begin{document}

\title{Outer Planet Single-Transit Detections with LSST}

\correspondingauthor{Derek Buzasi}
\email{dbuzasi@fgcu.edu}

\author[0000-0002-1988-143X]{Derek Buzasi}
\affil{Department of Chemistry and Physics, Florida Gulf Coast University, 10501 FGCU Blvd. S., Fort Myers, FL 33965 USA}



%


\section{} 
As of the end of January 2020, there were 4116 confirmed exoplanets\footnote{\url{https://exoplanetarchive.ipac.caltech.edu/}}, and that number will grow with
time as Kepler/K2 and TESS discoveries continue to be confirmed. Only $\sim$300 of
these planets have periods greater than 2 years, and none of
those have known densities derived using a combination of transit and radial velocity
measurements. 

I examine
the long-period single-transit detection rate expected using the Legacy Survey of Space and Time (LSST)\footnote{\url{https://www.lsst.org/lsst/}}. While the typical $\sim$10$\rm~deg^2$
LSST field will be observed roughly 1000 times over that period, there are four ``Deep Drilling Fields'' which will receive much more concentrated attention,
and thus are most likely to yield observable single transit events. Single-transit observations allow us to derive limits to orbital periods, and
knowledge of the location of the host stars and minimum transit depths will enable
ground-based follow-up using both photometric and radial velocity techniques \citep{2016MNRAS.457.2273O,2019AJ....157...84V}.

\begin{figure*}
  \centering
    \includegraphics[width=\textwidth]{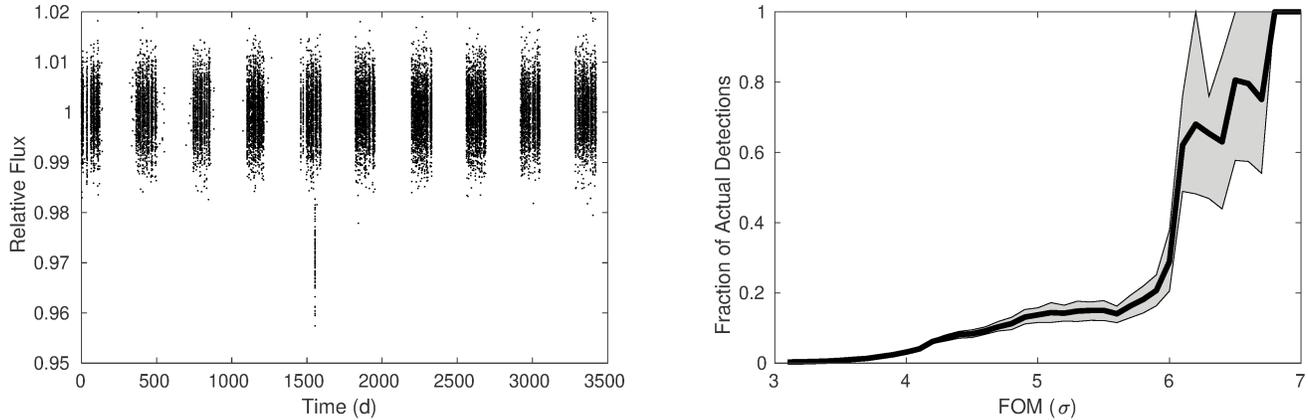}
  \caption{The left panel illustrates the light curve of a simulated example of a single transit event of a Jupiter-sized planet. The solid line in the right panel shows the frequency of actual detections as a function of the adopted figure of merit, while the shaded region illustrates $\pm 3 \sigma$ range of results from the complete simulation suite. Adopting ${\rm FOM} (\sigma) > 6$ reduces the false positive rate to under 50\%.  } \label{fig:detection}
\end{figure*}

I used the simulated LSST survey cadence from the Opsim v3.3.5
Minion 1016 Baseline Reference Survey \citep{2016SPIE.9910E..13D,2016SPIE.9911E..25R}.
I then simulated stellar populations using {TRILEGAL} \citep{2005A&A...436..895G,2009A&A...498...95V} to produce population models for the four pointings, with exoplanet occurrence rates from \cite{2016AJ....152..206F} over the ranges $2 < P < 25$ years and
$0.1 - 1 \rm R_J$. I constructed transit light curves based on the \cite{2002ApJ...580L.171M} formalism; for simplicity, I interpolated over quadratic limb-darkening coefficients from \cite{2010A&A...510A..21S}. I added noise based on the simulated characteristics of each individual observation according to the noise model of \cite{2019ApJ...873..111I} and generated light curves by combining simulations for all 6 colors, making the approximation that transits are color-independent (see Figure~\ref{fig:detection}).

I searched for single transit events using the algorithm of \cite{2012JASA....107..1590F} to search for changes in the mean value of the signal
lasting between 1 and 25 hours, adopting as figure of merit the mean
depth of the potential transit measured in standard deviations of the whitened original
light curve. To examine the influence of false positives, I ran similar simulations in each field without injected planets. 

The number of exoplanets detected at the $\rm FOM = 3\sigma$ level was $31 \pm 5$. However at this level, true detections are swamped by the roughly $3 \times 10^4$ false positives.
As illustrated in the right-hand panel of Figure~\ref{fig:detection}, increasing the FOM dramatically improves the ratio of real detections to false positives, although of course the number of detections also falls. Adopting
$\rm FOM = 6\sigma$ leads to $3.2 \pm 0.2$ detections and a similar number of false positives. This result is conservative due to the fairly primitive detection algorithm used and the disregard of any information from multiple transits and colors; one might reasonably anticipate several times as many detections if those factors are taken into account. The typical (median) detection is a $0.7~R_{\rm J}$ planet in a 4.8 year orbit around a late K dwarf with $g \sim 19.5$. Many of the predicted exoplanet detections should show detectable radial velocity variations using either existing spectrographs on 10m telescopes or planned ELT instruments \citep{2015arXiv150301770P}; 89\% show $\rm K > 10 ~m~ s^{-1}$. Most host stars are bright enough for ground-based transit follow-up using telescopes of modest
aperture, and the relatively compact fields in which they lie suggest that the use of dedicated instruments focused on these regions should be
explored, both in support of exoplanet science and follow-up of other transient and variable sources. Previous studies of LSST exoplanet detections have focused on detections of short-period systems \citep{2015AJ....149...16L,2015AJ....150...34J,2017AJ....153..186J}, but the long duration of the planned survey also lends itself to the detection of longer-period planets. While the number of potential detections is small, they would nevertheless represent a significant improvement on the current value.




\end{document}